\documentclass{article}
\usepackage[utf8]{inputenc}
\usepackage[margin=0.75in]{geometry}
\usepackage{authblk}
\usepackage[T1]{fontenc}
\usepackage[utf8]{inputenc}
\usepackage{changepage}
\usepackage{multicol}
\usepackage{blindtext}
\usepackage{floatrow}
\usepackage{caption}
\hyphenchar\font=-1
\title{Comparing Intellectual property policy in the Global North and South\\ - A one-size-fits-all policy for economic prosperity?}

\author[1]{S{.} Sidhartha Narayan}
\author[1]{Malavika Ranjan}
\author[1]{Madhumitha R}

\affil[1]{Indian Institute of Technology, Madras}

\date{}

\usepackage{graphicx}

\begin{document}

\maketitle

\begin{adjustwidth*}{+2cm}{+2cm} 
\begin{center}
   \emph{Abstract}
\end{center}

   \emph{This paper attempts to analyse policy making in the field of Intellectual Property (IP) as an instrument of economic growth across the Global North and South. It begins by studying the links between economic growth and IP, followed by an understanding of Intellectual Property Rights (IPR) development in the US, a leading proponent of robust IPR protection internationally. The next section compares the IPR in the Global North and South and undertakes an analysis of the diverse factors that result in these differences. The paper uses the case study of the Indian Pharmaceutical Industry to understand how IPR may differentially affect economies and conclude that there may not yet be a one size fits all policy for the adoption of Intellectual Property Rights.}
\end{adjustwidth*}

\vspace{1.5cm}

\section{Introduction}
\begin{multicols}{2}

Technological improvement was first introduced in Solow’s model as an exogenous variable in economic growth that was independent of the country’s economic policies. The endogenous growth theory, however, posits that government policies influence a country’s long-term growth in key areas such as intellectual property, taxation, maintenance of law and order, and fiscal and monetary policies (Idris 2002). It links faster pace of innovation and extra investment in human capital to improvement in productivity. Innovation results in technological transfer, positive spillovers, and revenue generation. This results in the improvement of Balance of Payments (BoP), access to international markets, creating confidence among investors, and increased employment opportunities. Hence, it is important for innovation-nurturing institutions, such as the government and the private sector, to incentivise inventiveness. \\
\\
IPRs are such incentives and contribute to economic growth through the channels of international trade, Foreign Direct Investments (FDI), licensing, and Research and Development (R$\&$D), i.e., innovation (Janjua and Samad 2007). A robust IPR regime facilitates the transfer of critical technology in the form of FDIs, joint ventures, and licensing. These channels help in providing access to technological and managerial assets of Multinational Companies (MNCs), hence boosting technological transfer, crucial for countries with low levels of innovation. The technical know-how induced by the IPRs expands business opportunities. Patent protection – laws that prevent inventions from being commercially made, used, distributed, imported or sold by others without the patent owner's consent – is a significant aspect of IPR as it promotes technological and business competitiveness. IPR protection measures also address the problem of counterfeit products, which cause massive annual losses to industries and reduce tax collections due to lack of documentation (Janjua and Samad 2007).  \\
\\
There is a high degree of variation in the relationship between IPR protection and economic growth. This variation is dependent upon the market structure and the nature of economic institutions of a country. It should be noted that countries with an open economic regime and a strong IPR protection system like the OECD countries have recorded some of the highest growth rates (Gould and Gruben 1997).

   \end{multicols}
   
\pagebreak   

\section{Development of IPR through the US and international cooperation}
\begin{multicols}{2}
The expansion of IPR, and the emergence of the US as a key champion of the world-wide adoption of IPR, is a result of strong legal foundations combined with increased levels of industrial and technological development. However, there is some difference of opinion even in this case. In the 1860s, patents in the US experienced a remarkable surge and played an important role in allowing electrical, chemical and communications industries to thrive. By the 19th century, instead of using the patents to just recoup inventions in specific industries, businesses strategically used patent rights for asserting and retaining their market control. This fear of monopolistic control over the market has been the major source of skepticism in the US regarding IPR, often expressed in tandem with anti-competition allegations.\\
\\
The adoption of IPR  gives great monetary benefit to inventors and their companies, so extensive lobbying by interest groups is commonplace. Those who would prefer weaker IPR enforcement tend to be limited and diluted consumers, who stand to lose out in the case of monopolistic markets. Thus, attempts to garner support for IPR protection have been largely successful in the US, and have been expanding since the 1970s, with increasing patents and claims filed in the rapidly developing software sector. (Bracha, 2017). As of 2021, the US has the best environment for intellectual property rights (Statista, 2021). A trend that is mirrored in several other Western countries like the UK, there has been increasing international support for the adoption and the enforcement of IPR. \\
\\
Noteworthy international conventions on intellectual property rights, that has caused a paradigm shift in world trade, are the ‘The World Intellectual Property Organization’ (WIPO of 1967) and the Trade-Related (Aspects of) Intellectual Property Rights (TRIPS) Agreement, which standardised a minimum level of IP protection among the member countries. Most developing countries are signatories and need to comply with the TRIPs agreement, explaining the basic level of  IPR protection in the Global South. However, the South is majorly slow-paced in the implementation of these IPR policies. Laggard implementation often leads to repercussions in access to the international market, withdrawal of the Generalized System of Preferences (GSP) and foreign investor confidence (Bracha, 2017). 

\end{multicols}

\section{Why does the Global South have Weak IPR Protection?}
\begin{multicols}{2}
Logically, the next question is – why are weaker IP Laws prevalent in the Global South? Is it more than just lousy policy-making? Yes. Bodies of research purport that economies of developing countries are not yet fertile for stringent IPR and might even be harmed by it (NERA, 2008). There is a significant variation in the effectiveness of IPR regimes between developed and developing countries. The coefficient associated with IPR with balanced data indicated that a one-unit increase in the IPR index caused a 0.73$\%$ decline in the real GDP per capita growth for middle-income developing countries (Janjua and Samad 2007). Being in a transitional stage of economic development, their cost of innovation is higher than the cost of imitation. Strong IPR measures run the risk of causing inflationary pressures, unemployment, and BoP problems at their current level of economic and infrastructural development as well (Gould and Gruben 1997). \\
\\
Stronger IPR protection may also contract FDI because the stronger uniqueness requirements increases the cost of production. The links between welfare and strong IPR remains ambiguous with studies showing that increasing IPR protection in the South had a negative welfare effect in the South and a positive effect in the North. Even a globally harmonised IPR would benefit the North but had the potential to harm the South. In the presence of evidence that suggests that there is no effect of IPR protection on growth for lower-income countries, the Global South is further deterred from adopting strong IPR protection. IPR protection only becomes important for countries as they reach higher levels of development because a major part of the benefits in the initial stages accrue to firms outside the home country. \\
\\
Nonetheless, an effectively managed, modern intellectual property system is a requisite for the present day economic development soaked in technology – a system which is not maintained in most of the developing countries. (NERA, 2008) The Global South faces other challenges such as weak physical infrastructure in terms of IP offices, low intellectual infrastructure, poor public awareness and a lack of foolproof government policies (Mashelkar 2002). R$\&$D institutions and firms in the Global South have modelled themselves to work around imitative research, reverse engineering and other non-original methods, hence bringing down the IP portfolio for the country. (Janjua, Zamurrad, Samad, 2007)

\end{multicols}

\pagebreak

\section{IPR Protection in the Global South: Other Factors of Influence}
\begin{multicols}{2}
A key concern that most developing countries have is that stronger IPR protection might enable foreign interests to reap the economic benefits of indigenous knowledge and other biological resources. One example of such a concern is “bioprospecting”. This is a practice by which foreign interests use biological samples to produce patentable products, which mostly build on indigenous knowledge. This may prevent people from benefiting from their generational cultural knowledge. It may be remedied by adding a measure of assuring that “traditional knowledge” is part of the collection consulted by patent examiners (Swiderska 2006). \\
\\
Though the connection between politics and IPR is tenuous, the effects of a strong IPR system are different under different regimes. For example, a study found that entrepreneurs in more democratic countries enjoy higher levels of technology usage as IPR strengthens whereas entrepreneurs in more autocratic countries are less likely to use the latest technology. The ability of early technology entrepreneurs to influence norms and laws to their advantage is greater with a higher degree of political participation. The prevalence of pirate parties – political interest groups who lobby for minimal IPR because they are of the view that it is against information and knowledge sharing – is also a signal that enforcement needs to occur in a way that does not stifle innovation. However, mechanisms such as corruption, nepotism, caste system might lead to a lopsided system where IPR is routinely enforced in favour of the politically connected (Laplume, Pathak and Xavier-Oliveria 2014). \\
\\
The setting up of strong IPR protection and sound enforcement mechanisms is also, to some degree, dependent on the culture and politics of each country. For example, though China is a member of the WIPO, the Paris Convention, the Berne Convention, and the Madrid Protocol, efficient enforcement mechanisms are a significant challenge. Some scholars argue that this is due to the commonly held belief in Chinese society that individual inventions draw on past knowledge, which belongs to all citizens (Alfors 1995). Confucianism and communism have historically placed more importance on curtailing unacceptable ideas as opposed to copyrights. Individual ownership is not a priority and inventions are understood to be State-owned (Zimmerman 2013). \\
\\
This line of reasoning is countered by the argument that formal institutions, the “rules of the game” and governance have a greater impact as opposed to informal institutions, which take a relatively longer time to change (Peng 2013). Olson’s argument about the quality of institutions and economic policies as important determinants of a country’s wealth (Olson 1996) becomes pertinent as a scholarship promoting an institution-based view of IPR protection is emerging. Weak institutions for IPR enforcement are characterised by a lack of checks and balances, concentration of power, high susceptibility to lobbying and political instability. This results in lower levels of domestic investment, and eventually capital flight (Svensson 1998).

\end{multicols}

\section{Case Study: Indian Pharmaceutical Industry and IPR}
\begin{multicols}{2}
In this case study, we establish that the growth of the Indian domestic pharmaceutical industry has  relied on weak instead of strong IPR, an important example to illustrate that one policy may not fit all. \\
\\
The Patents Act in 1970 played a historic role in weakening patent laws in the pharmaceuticals and agrochemical products space in India. The 1970 Patents Act excluded pharmaceuticals and agrochemical products from patent eligibility, in an attempt to reduce India's dependence on imports for bulk drugs and formulations. The goal was to develop a self-sufficient, indigenous pharmaceutical industry. The lack of protection for product patents had a significant impact on the Indian pharmaceutical industry. It resulted in the development of considerable expertise in reverse engineering of drugs – ones that are patentable as products throughout the industrialised world but unprotected in India. Consequently, the Indian pharmaceutical industry grew rapidly by developing cheaper versions of a number of drugs patented for the domestic market and eventually moved aggressively into the international market with generic drugs once the international patents expired. In addition, the Patents Act provides a number of safeguards to prevent abuse of patent rights and provide better access to drugs.\\
\\
The Indian pharmaceutical industry is a thriving, technologically-intensive industry that has witnessed consistent growth over the past three decades. Dominated by several privately owned Indian companies, growth has been facilitated by favourable government policies and limited overseas competition. However, after the liberalisation of the Indian economy in 1992, the Indian pharmaceutical industry is now being forced to revisit its long-term strategies and business models. With growing recognition of the need to ensure the protection of valuable investments in R$\&$D, factors such as IPR protection are becoming increasingly significant. In a growing list of efforts being made in the country against the weak enforceability of existing IP legislation, the Indian government is moving towards establishing a patent regime – one that is conducive to technological advances, and in keeping with its global commitments. A signatory to the TRIPS Agreement since 1994, India is required to meet minimum standards in relation to patents and the pharmaceutical industry. India's patent legislation must now include provisions for the availability of patents for both pharmaceutical products and process inventions. Patents are to be granted for a minimum term of 20 years to any invention of a pharmaceutical product or process that fulfils established criteria.

\end{multicols}

\section{Conclusion}
\begin{multicols}{2}
Though a causal relationship exists between IPR and economic prosperity, the benefits are not uniform across countries. There is still controversy surrounding the links between IPR protection and economic growth. Being in a transitional stage of development, costs of innovation are higher than imitation for developing countries. Imitating foreign technology is more profitable, and a robust IPR regime will hinder such developments. There is also evidence to suggest IPR protection impedes welfare in the Global South. Moreover, a country’s market structure also influences the link between IP rights, innovation, and growth.  \\
\\
Commonsensically, to join the trajectory of economic growth, the ideal step for the Global South should be to remodel their IP Laws on similar lines of the North and join. However, the two regions do not function independently. The IP laws of the Global North have an impact on the Global South’s policies and vice versa. Changing any policy will create a flux at the macro level and countries need to find the balance that suits their innovation and growth landscape. \\
\\
Thus, whether to strengthen IP Laws or not, remains an unanswered question. There exists no ‘one-size fits all’ policy for economic prosperity. Social, political, and demographic subjectivities also have to be taken into account while shaping the policy. As far as the stakeholders are considered, their benefits and losses depend on where the economy as a whole stands. While industries in the North might benefit from stringent IP policies, the same stakeholders in the South gain from looser rules of IP (since a majority rely on imitation or borrowed technology). Moreover, the time-frame also needs to be considered by the developing economies. For instance, looser IPR might be conducive for growth in the short-run (for imitation based stakeholders) but will be detrimental to innovation in the longer run for overall entrepreneurship. Eventually, economic prosperity is not a product of any one policy. Rather, it is a combination of efficient political institutions which shape policies best suited for an economy.

\end{multicols}

\section*{References}
%\begin{multicols}{2}
%\onehalfspacing
\noindent[1] Alford, William P. To steal a book is an elegant offense: Intellectual property law in Chinese civilization. Stanford University Press, 1995.
\\\\					 
\noindent[2] Baumol, William J., and Robert J. Strom. "Entrepreneurship and economic growth." Strategic entrepreneurship journal 1, no. 3‐4 (2007): 233-237.
\\\\
\noindent[3] Bracha, Oren. "The Emergence and Development of United States Intellectual Property Law." The Oxford Handbook of Intellectual Property Law.
\\\\
\noindent[4] Cimoli, Mario, Giovanni Dosi, Keith E. Maskus, Ruth L. Okediji, Jerome H. Reichman, and Joseph E. Stiglitz, eds. Intellectual property rights: legal and economic challenges for development. Oxford University Press, 2014.
\\\\
\noindent[5] Gould, David M., and William C. Gruben. 1997. “The Role of Intellectual Property Rights in Economic Growth.” Dynamics of Globalization and Development 59:209-241.
\\\\
\noindent[6]  Gupta, Anubhav. “Legal and ethical issues faced by start-ups in India”. Mondaq (2021).
\\\\					
\noindent[7] Idris, Kamil. 2002. Intellectual Property: A Power Tool for Economic Growth. N.p.: World Intellectual Property Organization.
\\\\
\noindent[8] Janjua, Pervez Z., and Ghulam Samad. 2007. “Intellectual Property Rights and Economic Growth: The Case of Middle Income Developing Countries.” The Pakistan Development Review 46, no. 4 (Winter): 711-722.
\\\\
\noindent[9] Laplume, Andre O., Saurav Pathak, and Emanuel Xavier-Oliveira. "The politics of intellectual property rights regimes: An empirical study of new technology use in entrepreneurship." Technovation 34, no. 12 (2014): 807-816.
\\\\
\noindent[10] Mashelkar, R. A. "Intellectual property rights and the Third World." Current Science (2001): 955-965.
\\\\
\noindent[11] NERA Economic Consulting. “Intellectual property rights in developing nations”. Mondaq (2008).
\\\\
\noindent[12] Olson, Mancur Jr. "Big bills left on the sidewalk: why some nations are rich, and others poor." (1996): 3-24.
\\\\
\noindent[13] Organization of Economic Cooperation and Development (OECD). “The economic effect of counterfeiting and piracy” (2007)
\\\\
\noindent[14] Park, Walter G. "North-South Models of Intellectual Property Rights: An Empirical Critique." Review of World Economics / Weltwirtschaftliches Archiv 148, no. 1(2012): 151-80. Accessed June 6, 2021.\\ http://www.jstor.org/stable/41485790.
\\\\
\noindent[15] Peng, Mike W. "An institution-based view of IPR protection." Business Horizons 56, no. 2 (2013): 135-139.
\\\\
\noindent[16] Schumpeter, Joseph A. The Theory of Economic Development: An Inquiry into Profits, Capita I, Credit, Interest, and the Business Cycle. Routledge, 2017.
\\\\
\noindent[17] Singh, Manoj K, Sharma, Himanshu. “Entrepreneurial dilemma in make in India”. Mondaq (2015)
\\\\
\noindent[18] Statista.com. ‘‘International Intellectual Property Rights Index, by country’’.\\ https://www.statista.com/statistics/257583/gipc-international-intellectual-property-index/
\\\\
\noindent[19] Svensson, Jakob. "Investment, property rights and political instability: Theory and evidence." European economic review 42, no. 7 (1998): 1317-1341.
\\\\
\noindent[20] Swiderska, Krystyna. "Protecting traditional knowledge: a framework based on customary laws and bio-cultural heritage." IIED: London, UK (2006).
\\\\
\noindent[21] Von Mises, Ludwig. "Human action: A treatise on economics." (1949).
\\\\
\noindent[22] WTO. n.d. “What are intellectual property rights?” WTO. https://www.wto.org/english/tratop$\_$e/trips$\_$e/intel1$\_$e.htm.
\\\\
\noindent[23] Yandle, Bruce. "Bootleggers and baptists-the education of a regulatory economists." Regulation 7 (1983): 12.
\\\\
\noindent[24] Zimmerman, Alan. "Contending with Chinese counterfeits: Culture, growth, and management responses." Business Horizons 56, no. 2 (2013): 141-148.

%\end{multicols}

\end{document}